\documentclass[a4paper,10pt]{article}
\usepackage[utf8x]{inputenc}

\title{ Chern-Simons-Matter Theory}
\author{ Mir Faizal \\
Mathematical Institute, University of Oxford
\\ Oxford
OX1 3LB, United Kingdom 
 }

\begin{document}

\maketitle

\begin{abstract}
In this paper we  will  deform a ABJ theory in $\mathcal{N}= 3$ harmonic superspace without breaking any  supersymmetry. 
We will analyse this ABJ theory and show that it retains the full $\mathcal{N}= 6$ supersymmetry. We will then analyse 
the gauge fixing and ghost terms for this model in various gauges. We will also analyse the corresponding 
BRST and anti-BRST symmetries of this model. 
\end{abstract}

\section{Introduction}

Chern-Simons theories are also important in 
 condensed matter physics due to their relevance in
  fractional quantum Hall  effect
\cite{a}-\cite{d}.  In fractional quantum Hall  effect the electrons are described as bosons in  combined external and statistical magnetic
fields. At special values of the filling fraction the statistical field cancels the external field,
in the mean field sense. The system at these values of the filling fraction is 
described as a gas of bosons feeling no net magnetic field. These bosons condense into a homogeneous ground state. 
 Thus, by coupling  Chern-Simons theory  to the fermions
 in two dimensions, fermions can be described as charged bosons carrying
an odd integer number of  flux quanta. 
Recently, supersymmetric generalisation of fractional quantum Hall effect have 
 been investigated \cite{a1}-\cite{qh12}. Furthermore, a relation between 
fractional quantum Hall  effect and noncommutative field theories has also been investigated \cite{nn}-\cite{nn1}. 
Thus,  the noncommutative 
 field theories  have interesting condensed matter applications. 
Thus, it will be interesting to analyse 
what effect the graviphoton deformation can have on the
properties of supersymmetric quantum Hall  systems. In fact, it will also be interesting to analyse 
the  theory dual to these deformations by using 
 $AdS/CMT$ correspondence \cite{adscmt}-\cite{adscmt1}. 
This is 
another  motivation for  studding Chern-Simons theories.
It may be noted that the holography of two dimensional conformal field theories is special because
in all higher dimensional examples, the propagating modes of a bulk gauge field are dual to a symmetry current in the boundary
theory. However, in two dimensional case,  the boundary currents are captured by topological terms in the bulk. 
It will also be interesting to analyse the gravity dual to the deformed ABJ theory. 

In this paper we will analyse the ABJ theory in harmonic superspace. 
 The harmonic superspace variable parameterize the coset $SU(2)/U(1)$ and are
 well suited for analysing theories with hight amount of supersymmetry. Thus, the
harmonic superspace has been used for studding theories with
  $\mathcal{N} =2$ supersymmetry in four dimensions   \cite{h1}-\cite{h2}. They have also been used for studding theories with   
 $\mathcal{N}= 3$ supersymmetry in three dimensions \cite{h21}-\cite{h4}. The  ABJM theory has also been analysed in  
harmonic superspace \cite{ahs}. ABJM  theory is a superconformal Chern-Simons-matter theory with manifest
 $\mathcal{N} =6$ supersymmetry   which is expected to get enhanced to 
 $\mathcal{N} =8$ supersymmetry \cite{sabjm}-\cite{sabjm1}. In this theory 
gauge fields are governed by the Chern-Simons 
action and the matter fields live in the bifundamental representation of the gauge group $U(N) \times U(N)$ \cite{1}-\cite{ 4}. 
 It  is thought to be a low energy description of $N$ M2-branes  
on $C^4/Z_k$ orbifold because it coincides with the BLG theory 
for the only known example of a Lie $3$-algebra \cite{BL1}-\cite{blG}. 
A generalization of the ABJM theory is called the ABJ theory \cite{5}-\cite{5a2}. In this theory 
the matter fields live in the bifundamental representation 
of gauge group $U(M) \times U(N)$ with $M \neq N$. 
This theory  also has $\mathcal{N} =6$ supersymmetry, but unlike the ABJM theory,  
non-planar corrections to the two-loop dilatation generator of ABJ theory
 mix states with positive and negative parity, and this mixing is proportional to $M − N$ \cite{5a}. The 
ABJ theory reduces to the ABJM theory when  $M =N$, and there is clearly no mixing for the ABJM theory.

In string theory the $NS$  backgrounds cause a noncommutative   deformation of the spacetime    
 \cite{sw}-\cite{co} and the $RR$ backgrounds causes a non-anticommutative 
deformation of the  Grassmann  coordinates which in-turn partially break the 
supersymmetry of the theory \cite{ov}-\cite{beta4}. However, 
   gravitino backgrounds cause a noncommutative deformation between the spacetime and Grassmann coordinates  
 \cite{bgn}-\cite{gp1}. 
The non-anticommutative deformation of harmonic superspace has already been analysed \cite{8l}-\cite{ 84}.
As M-theory is dual to type II string theory a deformation of the string theory side will also generate a deformation on the M-theory side. So,
in this paper we will thus analyse a deformation of the ABJ theory  caused by a non-vanishing 
commutator between the spacetime and Grassman coordinates. 
It will also be interesting to analyse the gravity dual of the deformed ABJ theory. 
 This motivates the study of deformation of the ABJ theory.

We will analyse  the BRST and the anti-BRST symmetries of this 
theory in various gauges. The BRST  and the anti-BRST symmetries occur for theories with a gauge degrees of freedom   \cite{abcd}. 
In Landau and Curci-Ferrari  gauges the BRST
and the anti-BRST transformations along with few other transformations  generate  a algebra known as the 
   Nakanishi-Ojima  algebra
\cite{n01}-\cite{n04}. In fact, this Nakanishi-Ojima algebra is mass-deformed in massive Curci-Ferrari  gauge \cite{n001}.
The BRST symmetry for the ordinary Chern-Simons theory 
\cite{16}-\cite{17} and $N = 1$ 
Chern-Simons theory \cite{18}-\cite{19} has been already studied. 
The BRST symmetry  of noncommutative pure Chern-Simons theory 
has also been analysed \cite{20g}-\cite{21g}. The BRST symmetry for the deformed ABJ theory has already been studied \cite{ref}. 
In this paper will  generalize this work to include the anti-BRST symmetry. 
Thus, we will analyse the BRST and the anti-BRST symmetries of the 
deformed ABJ theory in harmonic superspace.

\section{Harmonic superspace}
The $\mathcal{N} =3$ harmonic superspace  is constructed using the following derivatives 
\begin{eqnarray}
\nonumber \\ 
{\cal
D}^{++}&=&\partial^{++}+2i\theta^{++ a}\theta^{0 b}
 \partial^A_{ab}
 +\theta^{++a}\frac\partial{\partial\theta^{0 a}}
 +2\theta^{0 a}\frac\partial{\partial\theta^{--a}},\nonumber \\
{  D}^{--}&=&\partial^{--}
 -2i\theta^{--a}\theta^{0 b}\partial^A_{ab}
 +\theta^{--a}\frac\partial{\partial\theta^{0 a}}
 +2\theta^{0 a}\frac\partial{\partial\theta^{++ a}},
\nonumber \\
{  D}^0&=&\partial^0+2\theta^{++ a}\frac\partial{\partial\theta^{++ a}}
-2\theta^{--a}\frac\partial{\partial\theta^{--a}}, 
\end{eqnarray}
and
\begin{eqnarray}
D^{--}_a=\frac\partial{\partial\theta^{++ a}}
 +2i\theta^{--b}\partial^A_{ab}, &&
D^0_a= -\frac12\frac\partial{\partial\theta^{0 a}}
+i\theta^{0 b}\partial^A_{ab},\nonumber \\ 
D^{++}_a=\frac{\partial}{\partial
\theta^{--a}}, &&
\end{eqnarray}
where 
\begin{eqnarray}
&\partial^{++}=u^+_i\frac\partial{\partial u^-_i}, &
\partial^{--}=u^-_i\frac\partial{\partial u^+_i},\nonumber \\ 
&\partial^0 = u^+_i\frac\partial{\partial u^+_i}
-u^-_i\frac\partial{\partial u^-_i}.& 
\end{eqnarray}
Here the
harmonic variables $u^{\pm}$ are subjected to the 
constraints 
\begin{eqnarray}
 u^{+i} u^-_i = 1, && u^{+i} u^+_i = u^{-i} u^-_i =0. 
\end{eqnarray}
These derivatives are satisfy 
\begin{eqnarray}
 \{D^{++}_a, D^{--}_b\}=2i\partial^A_{ab}, \quad \{D^{0}_a,
D^{0}_b\}=-i\partial^A_{ab}, 
\nonumber \\ 
{[{ D}^{\mp\mp}, D^{\pm\pm}_a]}=2D^0_a, \quad [{  D}^{0},
D^{\pm\pm}_a]=\pm 2D^{\pm\pm}_a, 
\nonumber \\ 
\partial^0=[\partial^{++},\partial^{--}],\quad
[{  D}^{++}, {  D}^{--}]={  D}^0. \nonumber \\
\{D^{\pm\pm}_a, D^{0}_b\} = 0\,,\quad [{\cal
D}^{\pm\pm}, D^0_a]=D^{\pm\pm}_a.
  \end{eqnarray}
The full harmonic superspace is parameterized by 
\begin{equation}
 z = ( x^{ab}, \theta_{a}^{++}, \theta^{--}_a, \theta^0_a, u_i^{\pm} ),
\end{equation}
and the analytic
superspace 
is parametrized by
\begin{eqnarray}
\zeta_A=(x^{ab}_A,
\theta^{++}_a, \theta^{0}_a, u^\pm_i), 
  \end{eqnarray}
where
\begin{eqnarray}
x^{ab}_A=(\gamma_m)^{ab}x^m_A=x^{ab}
+i(\theta^{++a}\theta^{--b}+\theta^{++b}\theta^{--a}).
  \end{eqnarray}
This is because  the analytic
superspace is defined to be independent of the
$\theta^{--}_a$,
  \begin{eqnarray}
   D^{++}_a\Phi_A=0 \quad \Rightarrow \quad \Phi_A = \Phi_A(\zeta_A).
  \end{eqnarray}
In this superspace the generators of the supersymmetry are denoted by
\begin{eqnarray}
&Q^{++}_a=u^+_iu^+_j Q_a ^{ij}, &
Q^{--}_a=u^-_iu^-_j Q_a ^{ij},\nonumber \\ 
&Q^0_a = u^+_iu^-_j Q_a ^{ij},& 
\end{eqnarray}
where 
\begin{equation}
 Q_a^{ij} = \frac{\partial}{\partial \theta^a_{ij}} - \theta^{ijb }  \partial_{ab},
\end{equation}
and  the superspace measures are denoted by  
  \begin{eqnarray}
d^9z &=&-\frac1{16}d^3x
(D^{++})^2 (D^{--})^2(D^{0})^2, \nonumber \\
d\zeta^{(-4)}&=&\frac{1}{4} d^3x_Adu (D^{--})^2(D^{0})^2\,.
  \end{eqnarray}
A conjugation in this superspace is defined  by
  \begin{eqnarray}
\widetilde{(u^\pm_i)}=u^{\pm i},\quad \widetilde{(x^m_A)}=x^m_A, \nonumber \\  \widetilde{(\theta^{\pm\pm}_a)}=
\theta^{\pm\pm}_a,\quad \widetilde{(\theta^0_a)}=
\theta^0_a.
  \end{eqnarray}
Thus,   the analytic superspace measure is real
and the full superspace measure is imaginary. 

Now we can analyse the  deformation of this superspace, 
 caused by a gravitino background. This deformation gives rise to the following 
commutator 
\begin{equation}
 [\theta^{++a}, x^{\mu}] = A^{a \mu}, 
\end{equation}
and so this deformation does not break any  supersymmetry. 
This deformation induces the following star 
product in this superspace, 
\begin{eqnarray}
V^{++} (z) \star V^{++}  (z) &=&\exp -\frac{1}{2} \left(
A^{a \mu}( \partial^{2}_a \partial^{1}_\mu  + \partial^1_a \partial^2_\mu ) \right)
\nonumber \\ && \times 
 {V^{++}}(z_1) { V^{++}}  (z_2)
\left. \right|_{z_1=z_2=z}. 
\end{eqnarray}
Now we can construct the action for  ABJ theory in this deformed superspace using $V^{++}_L$ and $V^{++}_R$, which are defined by
\begin{eqnarray}
V^{++}_L &=& u^+_i u^+_j V^{ij}_L, \nonumber \\ 
V^{++}_R &=& u^+_i u^+_j V^{ij}_R,
\end{eqnarray}
where $V^{ij}_L$ and
$V^{ij}_R$ are  fields transforming under the gauge group  $U(M)$ and $U(N)$,  respectively. 
We also define matter fields $q^{+}$ and $\bar q^{+}$, which transform under the bifundamental representation of the group   $U(N) \times U(M)$.
 Now the action for this deformed ABJ theory, which is invariant under the gauge group $U(N) \times U(M)$,  can be written as 
\begin{equation}
 S = S_{CS, k} [ V^{++}_L]_ \star  + S_{CS, - k} [ V^{++}_R]_ \star   + S_{M} [ q^{+}, \bar q^{+}]_ \star,
\end{equation}
where 
\begin{eqnarray}
 S_{CS, k}[ V^{++}_L]_ \star &=&\frac{ik}{4\pi}\, tr\sum\limits^{\infty}_{n=2} \frac{(-1)^{n}}{n} \int
d^3x d^6\theta du_{1}\ldots du_n  H^{++}_L, \nonumber \\
  S_{CS, -k}[ V^{++}_R]_ \star &=&- \frac{ik}{4\pi}\,tr\sum\limits^{\infty}_{n=2} \frac{(-1)^{n}}{n} \int
d^3x d^6\theta du_{1}\ldots du_n H^{++}_R, \nonumber \\
S_{M} [ q^{+}, \bar q^{+}]_ \star &=&tr\int d\zeta^{(-4)}\bar q^{+} \star \nabla^{++}  \star q^{+},
\end{eqnarray}
and 
\begin{eqnarray}
 H^{++}_L &=& \frac{V^{++}(z,u_{1} )_L  \star V^{++}(z,u_{2} )_L\ldots
 \star V^{++}(z,u_n )_L }{ (u^+_{1} u^+_{2})\ldots (u^+_n u^+_{1} )},
\nonumber \\ 
H^{++}_R &=& \frac{V^{++}(z,u_{1} )_R  \star V^{++}(z,u_{2} )_R\ldots
 \star V^{++}(z,u_n )_R }{ (u^+_{1} u^+_{2})\ldots (u^+_n u^+_{1} )}, \nonumber \\ 
\nabla^{++}q^{+}&=&{  D}^{++}q^{+}
 + V^{++}_L \star q^{+}- q^{+} \star V^{++}_R, 
 \nonumber \\   \nabla^{++}\bar q^{+}&=&{  D}^{++}\bar q^{+}
 -\bar q^{+}  \star V^{++}_L +  V^{++}_R  \star \bar q^{+}.
\end{eqnarray}
The covariant derivatives for the matter fields in the deformed ABJ theory are given by  
\begin{eqnarray}
\nabla^{++}q^{+}&=&{  D}^{++}q^{+}
 + V^{++}_L \star q^{+}- q^{+} \star V^{++}_R\,,  \nonumber \\   \nabla^{++}\bar q^{+}&=&{  D}^{++}\bar q^{+}
 -\bar q^{+}  \star V^{++}_L +  V^{++}_R  \star \bar q^{+},
\end{eqnarray}
It is useful to define $V^{--}_L$ and $ V^{--}_R$ as
\begin{eqnarray}
 V^{--}_L&=&\sum_{n=1}^\infty (-1)^n \int du_1\ldots
du_n  E^{++}_L, \nonumber \\ 
 V^{--}_R&=& \sum_{n=1}^\infty (-1)^n \int du_1\ldots
du_n  E^{++}_R,
\end{eqnarray}
where 
\begin{eqnarray}
  E^{++}_L &=& \frac{V^{++}_L(z,u_1)\star V^{++}_L(z,u_2)\ldots \star 
V^{++}_L(z,u_n)}{(u^+u^+_1)(u^+_1u^+_2)\ldots (u^+_n u^+)}, \nonumber \\
 E^{++}_R &=& \frac{V^{++}_R (z,u_1)\star V^{++}_R(z,u_2)\ldots \star 
V^{++}_R(z,u_n)}{(u^+u^+_1)(u^+_1u^+_2)\ldots (u^+_n u^+)}.
\end{eqnarray}
It is also useful to define $ W^{++}_L$ and $ W^{++}_R $ as 
\begin{eqnarray}
  W^{++}_L &=& -\frac{1}{4} D^{++a} D^{++}_{ a}  V^{--}_L, \nonumber \\ 
 W^{++}_R &=& -\frac{1}{4} D^{++a} D^{++}_{ a}  V^{--}_R. 
\end{eqnarray}
This ABJ theory is invariant under the following infinitesimal gauge transformations
\begin{eqnarray}
\delta q^{+} &=& \Lambda_L  \star q^{+}-q^{+} \star \Lambda_R,\nonumber \\
 \delta\bar q^{+} &=&\Lambda_R  \star \bar q^{+}-\bar q^{+} \star \Lambda_L,\nonumber \\
\delta V^{++}_L&=&\nabla^{++} \Lambda_L,\nonumber \\
\delta V^{++}_R&=&\nabla^{++} \Lambda_R,   \label{1}
\end{eqnarray}
where 
\begin{eqnarray}
\nabla^{++} \star \Lambda_L&=&-{  D}^{++}\Lambda_L -[V^{++}_L,\Lambda_L]_ \star,\nonumber \\
\nabla^{++} \star \Lambda_R&=&-{  D}^{++}\Lambda_R -[V^{++}_R,\Lambda_R]_ \star,   
\end{eqnarray}
and the following $\mathcal{N}= 3$ supersymmetric transformations 
\begin{eqnarray}
\delta_\epsilon q^{+}&=& i\epsilon^{a}\hat\nabla^0_a \star q^{+}\,, \nonumber \\
\delta_\epsilon\bar q^{+} &=&i\epsilon^{a} \hat\nabla^0_a \star \bar q^{+ }\,, \nonumber  \\
\delta_\epsilon V^{++}_L&=&\frac{8\pi}k\epsilon^{a}
 \theta^0_a \star q^+\bar\star q^+\,, \nonumber \\
\delta_\epsilon V^{++}_R &=&\frac{8\pi}k\epsilon^{a}
 \theta^0_a \star \bar q^+ \star q^+\,,
\label{epsilon4}
\end{eqnarray}
where
\begin{eqnarray}
 \hat\nabla^0_a \star q^{+}& =& \nabla^0_a \star q^{+} 
+\theta^{--}_a (W^{++}_L \star q^{+} -q^{+} \star W^{++}_R )\,, \nonumber \\
 \nabla^0_a \star q^+&=&D^0_a q^+
 +V^0_{L\, a}\star q^+ -q^+\star V^0_{R\,a }\,, \nonumber \\ V^0_{L,R\, a}&=&-\frac12D^{++}_a
V^{--}_{L,R}. \nonumber \\ 
\end{eqnarray}
Thus, apart from the original manifest $\mathcal{N} =3$ supersymmetry, this model has additional $\mathcal{N} =3$ supersymmetry. 
So, this ABJM theory has $\mathcal{N} =6$ supersymmetry.

\section{Linear Gauge}
As the  deformed ABJ theory is invariant under gauge transformations  given by Eq. (\ref{1}), we can not quantize it without fixing 
a gauge. Thus, we  choose the gauge fixing conditions,
\begin{eqnarray}
D^{++} \star   V^{++}_L  =0, && D^{++}  \star V^{++}_R   =0. 
\end{eqnarray}
To incorporate these gauge fixing conditions at a quantum level, we add the following gauge fixing term  to 
the original Lagrangian density,
\begin{eqnarray}
\mathcal{L}_{gf} &=& \int d\zeta^{(-4)}  tr  \left[b_L \star (D^{++}   V^{++}_L ) + \frac{\alpha}{2}b_L \star b_L \right. \nonumber \\ && \left.  
\,\,\,\,\,\,\,\,\,\,\, -
 b_R  \star (D^{++} V^{++}_R  ) + \frac{\alpha}{2}b _R  \star  b_R 
\right]_|.
\end{eqnarray}
In order to ensure unitarity of the model we also add the following ghost term  to 
the original Lagrangian density, 
\begin{equation}
\mathcal{L}_{gh} = tr\int d\zeta^{(-4)} 
[ \overline{c}_L  \star D^{++} \nabla^{++}  \star c_L - \overline{c}_R  \star D^{++} \nabla_a \star c_R ]_|.
\end{equation}
The  sum of the gauge fixing term  and the ghost term, 
$\mathcal{L}_{g} = \mathcal{L}_{gf} + \mathcal{L}_{gh} $, is  a total BRST of $\Phi$, and a total anti-BRST variation of $\overline{\Phi}$, 
\begin{eqnarray}
 \mathcal{L}_{g} &=&   \int d\zeta^{(-4)}   s\, tr  [\Phi]_|\nonumber \\
 &=& - \int d\zeta^{(-4)}  \overline{s}\, tr [\overline{\Phi}]_|,
\end{eqnarray}
where 
\begin{eqnarray}
\Phi &=&  c_L \star \left(D^{++}   V^{++} _L   
 -  \frac{i\alpha}{2}b_L \right) 
-  c_R \star \left(D^{++}   V^{++}_R    
 -  \frac{i\alpha}{2}b_R\right),  \nonumber\\
 \overline{\Phi} &=&  \overline{c}_L \star \left(D^{++}    V^{++} _L 
 -  \frac{ \alpha}{2}b_L\right) 
-   {\overline{c}}_R \star \left(D^{++}    V^{++}_R   
 -  \frac{ \alpha}{2} b_R\right).
\end{eqnarray}
Here the BRST transformations are given by 
\begin{eqnarray}
s \,V^{++}_L = \nabla^{++} \star  c_L, && s\, V^{++}_R =\nabla^{++}_R  \star   c_R, \nonumber \\
s \,c_L = - {[c_L,c_L]}_ {\star} , && s \,\overline{c}_R =-  b_R - 2 [\overline{c}_R ,  c_R]_{\star}, \nonumber \\
s \, \overline{c}_L = b_L, && s \, c_R = - [c_R ,  c_R]_{\star}, \nonumber \\ 
s \,b_L =0, &&s \, b_R= - [ b_R, \overline{c}_R]_{\star}, \nonumber \\ 
s\, q^{+} = c_L  \star q^{+}-q^{+} \star c_R,&&
s\, \bar q^{+} = c_R  \star \bar q^{+}-\bar q^{+} \star c_L,
\end{eqnarray}
and the anti-BRST transformations are given by
\begin{eqnarray}
\overline{s} \,V^{++} = \nabla^{++}  \star \overline{c}_L, &&  \overline{s} 
\,  V^{++}_R =  \nabla^{++} \star \overline{c}_R,\nonumber \\
\overline{s} \,c_L = -b_L - 2 [\overline{c}_L, c_L]_ \star,  && \overline{s} \, c_R =   b_R, \nonumber \\
\overline{s} \,\overline{c}_L = - [\overline{c}_L, \overline{c}_L]_\star, 
                   &&\overline{s} \,{\overline{c}}_R = -[ {\overline{c}}_R,{\overline{c}}_R]_\star,\nonumber \\ 
\overline{s} \,b_L =- {[b_L,c_L]}_\star,  && \overline{s}  \,b_R  = 0,
 \nonumber \\ 
\overline{s} \, q^{+} = \overline{c}_L  \star q^{+}-q^{+} \star \overline{c}_R,&&
\overline{s} \, \bar q^{+} = \overline{c}_R  \star \bar q^{+}-\bar q^{+} \star \overline{c}_L.
\end{eqnarray}
Both these transformations are nilpotent,
$s^2 =  \overline s^2 =0 $. In fact, they also satisfy, $ s\overline s + \overline s s = 0$. Now as the sum of the 
ghost term and the gauge fixing term is  expressed 
as a total BRST or a total anti-BRST variation, it invariant under them. This is   because 
of the nilpotency of these  transformations. The 
BRST or the anti-BRST variation of the original classical Lagrangian density  are its gauge variations with the gauge parameter 
replaced by the ghosts or the anti-ghosts. As the original classical Lagrangian density was gauge invariant, so it is also  
invariant under these BRST and anti-BRST transformations. 
Thus, the effective Lagrangian density, which is defined to be a sum of the original classical Lagrangian density, the gauge 
fixing term and the ghost term, is also invariant under  these BRST and anti-BRST transformations. 
The  sum of the gauge fixing term and ghost term takes a simple form in the Landau gauge, $\alpha = 0 $,  
\begin{eqnarray}
\mathcal{L}_{g}  &=&\int d\zeta^{(-4)}   s\,   tr \left[ \overline{c}_L \star (D^{++}   V^{++} _L  )  
- {\overline{c}_R} \star(D^{++}  V^{++}  _R )\right]_| \nonumber \\ 
&=& \int d\zeta^{(-4)} \overline{s}\,  tr \left[ c_L \star (D^{++}   V^{++} _L   ) - c_R \star(D^{++}   V^{++} _R   )  \right]_|.
\end{eqnarray}
In fact, in  Landau gauge this can be expressed as combination of a total BRST and a total anti-BRST variation. Thus,  in Landau gauge
  sum of the gauge fixing term and the ghost term is given by 
\begin{eqnarray}
\mathcal{L}_{g} &=& -\frac{1}{2}\int d\zeta^{(-4)} s  \overline{s}\,tr [  \mathcal{Z}  ]_| \nonumber \\ 
&=& \frac{1}{2}\int d\zeta^{(-4)}   \overline{s} s \, tr [  \mathcal{Z} ]_|,
\end{eqnarray}
where 
\begin{equation}
 \mathcal{Z}  = V^{++}_L    \star   V^{++}_L  - 
  V^{++}    _R \star  V^{++}_R.
\end{equation}

\section{ Non-Linear Gauges}
For gauge theories sum of the gauge fixing term and the ghost term can also be expressed 
as a combination of the total BRST and the total anti-BRST variation, 
 for any value of $\alpha$ in Curci-Ferrari gauge \cite{n01}-\cite{n04}. 
Here we will show that   this also holds for a deformed ABJ theory  
in $\mathcal{N} = 3$ harmonic superspace formalism. 
The non-linear BRST transformations  for the deformed ABJ theory  in $\mathcal{N} = 3$ harmonic superspace are given by  
\begin{eqnarray}
s \,  V^{++}_L  = \nabla^{++} \star c_L, &&
s \,b_L  = - [b_L, c_L]_\star - [ \overline{c}_L, [ c_L,  c_L]_\star]_\star,\nonumber \\
s \,c_L  = - [c_L,  c_L]_\star,&&
s \,\overline{c}_L  = b_L - [\overline{c}_L, c_L]_\star, \nonumber \\ 
s \, V^{++}_R  =  \nabla^{++}_R \star c_R, &&
s \, b_R  = - {[b_R,  c_R]}_\star - [ {\overline{c}}_R, [  c_R, c_R]_\star]_\star \nonumber \\ 
s \, c_R = - [ c_R,   c_R]_\star, &&
s \,{\overline{c}}_R = b_R - [{\overline{c}}_R, c_R]_\star, \nonumber \\ 
s\, q^{+} = c_L  \star q^{+}-q^{+} \star c_R, &&
s\, \bar q^{+}= c_R  \star \bar q^{+}-\bar q^{+} \star c_L,
\end{eqnarray}
and the non-linear anti-BRST transformations  for the deformed ABJM theory  in $\mathcal{N} = 3$ harmonic superspace are given by  
\begin{eqnarray}
\overline{s}\,   V^{++}_L   = \nabla^{++} \star\overline{c}_L, &&
\overline{s} \,b_L  = - [b_L, \overline{c}_L]_\star +  [c_L,[\overline{c}_L,\overline{c}_L]_\star]_\star,\nonumber \\
\overline{s} \,\overline{c}_L  = - [\overline{c}_L , \overline{c}_L]_\star,&&
\overline{s} \,c_L  = - b_L - [\overline{c}_L, c_L]_\star, \nonumber \\ 
\overline{s}\,    V^{++}_R   =\nabla^{++} \star {\overline{c}}_R, &&
\overline{s} \,b_R = - [ b_R ,{\overline{c}}_R]_\star +  [ c_R ,[{\overline{c}}_R , {\overline{c}}_R]_\star]_\star,
\nonumber \\ 
\overline{s} \,{\overline{c}}_R  = - [{\overline{c}}_R , {\overline{c}}_R]_\star, &&
\overline{s} \,c _R =-  b_R - [{\overline{c}}_R,  c_R]_\star, \nonumber \\ 
\overline{s} \, q^{+} = \overline{c}_L  \star q^{+}-q^{+} \star \overline{c}_R,&&
\overline{s} \, \bar q^{+} = \overline{c}_R  \star \bar q^{+}-\bar q^{+} \star \overline{c}_L.
\end{eqnarray}
These transformations also satisfy $s^2 =  \overline s^2 = s\overline s + \overline s s=0 $. 
Thus, both these transformations are also nilpotent. 
We can now write sum of the gauge fixing term and the ghost term for this deformed ABJ  theory
 as a combination of a total BRST and a total  anti-BRST variation
\begin{eqnarray}
 \mathcal{L}_{g} &=& \frac{1}{2}\int d\zeta^{(-4)} s\overline{s} \, tr [\mathcal{Z} + \mathcal{Y}]_| \nonumber \\ 
 &=& - \frac{1}{2}\int d\zeta^{(-4)} \overline{s} s \, tr [\mathcal{Z} + \mathcal{Y}]_|,  
\end{eqnarray}
where 
\begin{equation}
 \mathcal{Y} = \alpha {\overline{c}}_R \star {c}_R - \alpha \overline{c}_L \star c_L.
\end{equation}

In gauge theories \cite{n01}-\cite{n04}, 
 the addition of a  bare mass term  breaks the nilpotency of the BRST and the anti-BRST transformations. 
Here we will show that   this also occurs for a deformed ABJ theory  
in $\mathcal{N} = 3$ harmonic superspace formalism.  The bare mass term is added to Curci-Ferrari model 
to obtain a massive Curci-Ferrari model as follows
 \begin{eqnarray}
\mathcal{L}_{g}&=& -\frac{1}{2}\int d\zeta^{(-4)} [\overline{s}s+im^2]\, tr \left[ \mathcal{Z} + \mathcal{Y} \right]_| \nonumber \\ 
&=& \frac{1}{2}\int d\zeta^{(-4)} [s\overline{s} - im^2]\, tr \left[  \mathcal{Z} + \mathcal{Y} \right]_|.
\end{eqnarray}
Now the BRST transformations get modified as follows   
\begin{eqnarray}
s \,  V^{++}_L   = \nabla^{++} \star c_L, &&
s \,b_L  = im^2 c_L - [b_L, c_L]_\star - [ \overline{c}_L, [ c_L,  c_L]_\star]_\star,\nonumber \\
s \,c _L = - [c _L ,c_L]_\star, &&
s \,\overline{c} _L = b_L - [\overline{c}_L, c_L]_\star, \nonumber \\ 
s \,  V^{++}_R  =   \nabla^{++} \star c_R, &&
s \, b_R  =im^2 {c}_R- {[ b_R,  c_R]}_\star - [ \overline{c}_R, [ c_R, c_R]_\star]_\star \nonumber \\ 
s \, c_R =- [ c_R,   c_R]_\star, &&
s \,{\overline{c}}_R = b_R - [{\overline{c}}_R,  c_R]_\star, \nonumber \\ 
\overline{s} \, q^{+} = \overline{c}_L  \star q^{+}-q^{+} \star \overline{c}_R,&&
\overline{s} \, \bar q^{+} = \overline{c}_R  \star \bar q^{+}-\bar q^{+} \star \overline{c}_L, 
 \end{eqnarray}
and  the anti-BRST transformations get modified as 
\begin{eqnarray}
\overline{s}\,   V^{++}_L   = \nabla^{++} \star\overline{c}_L,&&
\overline{s} \,b_L  = im^2 \overline{c}_L - [b_L, \overline{c}_L]_\star +  [c_L,[\overline{c}_L,\overline{c}_L]_\star]_\star,\nonumber \\
\overline{s} \,\overline{c}_L  = - [\overline{c}_L , \overline{c}_L]_\star, &&
\overline{s} \,c_L = - b_L - [\overline{c}_L, c_L]_\star, \nonumber \\ 
\overline{s}\,   V^{++} _R  =\nabla^{++} \star \overline{c}_R, &&
\overline{s} \,b _R = im^2 {\overline{c}} _R - [b_R, {\overline{c}}_R]_\star + 
                           [ c_R ,[{\overline{c}}_R,{\overline{c}}_R]_\star]_\star,
                          \nonumber \\ 
\overline{s} \,\overline{c}_R   = - [{\overline{c}}_R ,{\overline{c}}_R]_\star, &&
\overline{s} \, c _R = -  b _R - [{\overline{c}}_R, c_R]_\star, \nonumber \\ 
\overline{s} \, q^{+} = \overline{c}_L  \star q^{+}-q^{+} \star \overline{c}_R,&&
\overline{s} \, \bar q^{+} = \overline{c}_R  \star \bar q^{+}-\bar q^{+} \star \overline{c}_L.
\end{eqnarray}
These modified  BRST and  anti-BRST transformations now satisfy
\begin{equation}
s^2 = \overline{s}^2  \sim 2i m^2.
\end{equation}
Thus, the addition of the bare mass term breaks the nilpotency of the BRST and the anti-BRST transformations.  
However,  in the zero mass limit, 
the nilpotency of the BRST and the anti-BRST transformations is restored. 

\section{Nakanishi-Ojima Algebra}
When ever the sum of the  
gauge fixing term and the ghost term can be written as 
a combination of the total BRST and the total anti-BRST variation, 
the total Lagrangian density is invariant under  a set of symmetry 
transformations which obey a
$SL(2, R)$ algebra called  Nakanishi-Ojima algebra. 
We will show that this algebra also hold for the ABJ theory in $\mathcal{N} =3$ harmonic superspace. To do so we first 
define the following transformations 
for the deformed   ABJ 
theory, 
\begin{eqnarray}
 \delta_{1}\, b_L =  [ c_L,  c_L]_\star, 
& \delta_{1}\,  b _R = 
  [  c_R, c_R ]_\star, 
& \delta_{1}\, c_L = 0, 
\nonumber \\
  \delta_{1}\,  c_R =  0,
  &\delta_{1}\, \overline{c}_L =  c_L,
 & \delta_{1}\,  {\overline{c}}_R =  {c}_R, 
\nonumber \\
 \delta_{1}\,   V^{++}_L  = 0, 
 & \delta_{1}\,  V^{++}_R   =   0,
 & \delta_{1}\, q^+ = 0, 
\nonumber \\ 
  \delta_{1}\,  \bar q^+ =   0,
& \delta_{2}\, b_L =   [ \overline c_L,  \overline c_L]_\star, 
  &\delta_{2}\,  b_R=   
[ {\overline{c}}_R, {\overline{c}}_R]_\star, 
\nonumber \\
 \delta_{2}\, c_L =  \overline{c}_L, 
&  \delta_{2}\,  c_R = 
{\overline{c}}_R, 
 & \delta_{2}\, \overline{c}_L =  0,
\nonumber \\ 
  \delta_{2}\,  {\overline{c}}_R =  0,
& \delta_{2}\, q^+ =  0  
& \delta_{2}\,  \bar q^+ =   0.
\end{eqnarray}
Now we can see that in Landau and Curci-Ferrari gauges these 
 transformations, the BRST transformation and the anti-BRST transformation 
 along with the $FP$-conjugation  form the 
Nakanishi-Ojima $SL(2, R)$ algebra,
\begin{eqnarray}
 [s,s]_{\star} =0, && [\overline{s},\overline{s}]_{\star} =0, \nonumber \\ 
{[s, \overline{s}]}_\star =0, && [\delta_{1}, \delta_{2}]_\star = - 2 \delta_{FP} \nonumber \\ 
{[\delta_{1}, \delta_{FP}]}_\star = -4 \delta_{1}, && [\delta_{2}, \delta_{FP}]_\star = 4 \delta_{2},\nonumber \\ 
{[s, \delta_{FP}]}_\star = - 2s , && [\overline s, \delta_{FP}]_\star = 2\overline s,\nonumber \\ 
{[s, \delta_{1}]}_\star = 0, && [\overline{s}, \delta_{1}]_\star = -2 s,\nonumber \\ 
{[s, \delta_{2}]}_\star = 2 \overline{s}, && [\overline{s}, \delta_{2}]_\star = 0.
\end{eqnarray}
A bare mass term breaks  the nilpotency of the BRST and the anti-BRST transformations in the massive  Curci-Ferrari gauge. However,  the 
$FP$-conjugation is not broken in the massive  Curci-Ferrari gauge. Thus, we are  able construct a mass-deformed 
version of this Nakanishi-Ojima  in massive Curci-Ferrari gauge,
\begin{eqnarray}
 [s,s]_{\star} =-2im^2 \delta_{1}, && [\overline{s},\overline{s}]_{\star} =2im^2 \delta_{2}, \nonumber \\ 
{[s, \overline{s}]}_\star =2im^2 \delta_{FP}, && [\delta_{1}, \delta_{2}]_\star = - 2 \delta_{FP} \nonumber \\ 
{[\delta_{1}, \delta_{FP}]}_\star = -4 \delta_{1}, && [\delta_{2}, \delta_{FP}]_\star = 4 \delta_{2},\nonumber \\ 
{[s, \delta_{FP}]}_\star = - 2s , && [\overline s, \delta_{FP}]_\star = 2\overline s,\nonumber \\ 
{[s, \delta_{1}]}_\star = 0, && [\overline{s}, \delta_{1}]_\star = -2 s,\nonumber \\ 
{[s, \delta_{2}]}_\star = 2 \overline{s}, && [\overline{s}, \delta_{2}]_\star = 0.
\end{eqnarray}
\section{Physical States}
It is now possible to construct  current corresponding to the BRST and the anti-BRST symmetries 
of this theory. 
The current 
associated with the   BRST symmetry  is given by
\begin{eqnarray}
2J_{(B)}^\mu & = &\int d\zeta^{(-4)}  tr 
\left[ \frac{ \partial L_{eff}  }{\partial \mathcal{D}_{\mu} V^{++}_L  } 
\star  s\, 
V^{++}_L  +
 \frac{ \partial L_{eff}  }{\partial  \mathcal{D}_{\mu} c_L } \star  s\, c_L
   +
\frac{ \partial L_{eff}  }{\partial  \mathcal{D}_{\mu} \overline{c}_L }
 \star  s\, \overline{c}_L \right. \nonumber \\&&
 \,\,\,\,\,\, \,\,\,\,\,\,\, \,\,\,\,\,\, \,\,\,\,\,\,\,\,\,\, +
\frac{ \partial L_{eff}  }{\partial \mathcal{D}_{\mu} b_L } \star   s\, b_L
 +
\frac{ \partial L_{eff}  }{\partial \mathcal{D}_{\mu}  V^{++}_R } \star  s\, 
V^{++} _R +
 \frac{ \partial L_{eff}  }{\partial  \mathcal{D}_{\mu}  c_R }
 \star  s\, ]
c_R
 \nonumber \\&&
 \,\,\,\,\,\, \,\,\,\,\,\,\, \,\,\,\,\,\, \,\,\,\,\,\,\,\,\,\, +
\frac{ \partial L_{eff}  }{\partial  \mathcal{D}_{\mu}{\overline{c}}_R }
 \star  s\, \overline{c}_R + 
\frac{ \partial L_{eff}  }{\partial \mathcal{D}_{\mu}  b_R } \star  
 s\,
  b_R +
\frac{ \partial L_{eff}  }{\partial \mathcal{D}_{\mu}  q^+ } \star  
 s\,
 q^+
 \nonumber \\&&\left. 
 \,\,\,\,\,\, \,\,\,\,\,\,\, \,\,\,\,\,\, \,\,\,\,\,\,\,\,+ 
\frac{ \partial L_{eff}  }{\partial \mathcal{D}_{\mu}  \bar{q}^+ } \star  
 s\,
 \bar{q}^+
 \right]_|,
\end{eqnarray}
and current 
associated with the
 anti-BRST symmetry is given by
\begin{eqnarray}
\nonumber \\
 2\overline{J}_{(B)}^\mu  & = &\int d\zeta^{(-4)}  tr 
\left[ \frac{ \partial L_{eff}  }{\partial \mathcal{D}_{\mu} V^{++}_L  }
 \star \overline{s}\, 
V^{++}_L +
 \frac{ \partial L_{eff}  }{\partial  \mathcal{D}_{\mu} c_L } \star  \overline{s}\, c_L
  +
\frac{ \partial L_{eff}  }{\partial  \mathcal{D}_{\mu} \overline{c}_L }
 \star  \overline{s}\, \overline{c}_L \right. \nonumber \\&& 
 \,\,\,\,\,\, \,\,\,\,\,\,\, \,\,\,\,\,\, \,\,\,\,\,\,\,\,\,\,+ 
\frac{ \partial L_{eff}  }{\partial \mathcal{D}_{\mu} b_L } \star 
  \overline{s}\, b_L
 +
\frac{ \partial L_{eff}  }{\partial \mathcal{D}_{\mu}  V^{++}_R } 
\star  \overline{s}\, 
V^{++}_R  +
 \frac{ \partial L_{eff}  }{\partial  \mathcal{D}_{\mu}c_R } \star  \overline{s}
\, 
c _R
 \nonumber \\&&
 \,\,\,\,\,\, \,\,\,\,\,\,\, \,\,\,\,\,\, \,\,\,\,\,\,\,\,+
\frac{ \partial L_{eff}  }{\partial  \mathcal{D}_{\mu} {\overline{c}}_R }
 \star  \overline{s}\, {\overline{c}}_R + 
\frac{ \partial L_{eff}  }{\partial \mathcal{D}_{\mu}  b _R} \star  
 \overline{s} \,
  b_R
+
\frac{ \partial L_{eff}  }{\partial \mathcal{D}_{\mu}  q^+ } \star  
  \overline{s}\,
 q^+
 \nonumber \\&&\left. 
 \,\,\,\,\,\, \,\,\,\,\,\,\, \,\,\,\,\,\, \,\,\,\,\,\,\,\,\,\, + 
\frac{ \partial L_{eff}  }{\partial \mathcal{D}_{\mu}  \bar{q}^+ } \star  
  \overline{s}\,
 \bar{q}^+
\right]_|,
\end{eqnarray}
where
\begin{equation}
 \int d\zeta^{(-4)}  [L_{eff}]_| = \mathcal{L}_c + \mathcal{L}_{gh}
 + \mathcal{L}_{gf}.
\end{equation}
 If we restrict the deformations to space-like deformations, 
then the  charges corresponding to the BRST and the anti-BRST symmetries  are given by 
\begin{eqnarray}
 Q_B &=& \int d^3 y  \, \,  J_{(B)}^0, \nonumber \\ 
 \overline{Q}_B &=& \int d^3 y  \, \,  \overline{J}_{(B)}^0.
\end{eqnarray}
So, from now on we shall  
  restrict the deformations to
space-like deformations. Now  these  charge associated with
the BRST and the anti-BRST symmetries transformation commutes
 with the
Hamiltonian and  are  conserved. These charges 
 are nilpotent for all gauges except the massive Curci-Ferrari gauge, 
\begin{equation}
 Q_B^2 =  \overline{Q}_B^2 =0. 
\end{equation}
However, for massive Curci-Ferrari gauge these charges are not nilpotent 
\begin{eqnarray}
  Q_B^2 &\neq& 0, \nonumber \\
\overline{Q}_B^2 &\neq& 0. 
\end{eqnarray}
 Now we define physical states as states that are annihilated by $Q_B$
\begin{equation}
 Q_B |\phi_p \rangle =0. 
\end{equation}
We can equivalently  define the physical states as states that are annihilated by $\overline{Q}_B$
 \begin{equation}
 \overline{Q}_B |\phi_p \rangle =0. 
\end{equation}
The inner product of those
 physical states, which are obtained from unphysical states by the action of these charges, 
 vanishes with all other physical states. This is because if $|\phi_{up}\rangle$ is a unphysical state, then 
\begin{eqnarray}
 \langle \phi_p|Q_B|\phi_{up}\rangle&=&0, \nonumber \\ 
  \langle \phi_p|\overline{Q}|\phi_{up}\rangle&=&0.
\end{eqnarray}
So, all the relevant physical information lies in the physical states which are not obtained by the action of these 
charges on unphysical states. 
 Now if the asymptotic physical states are given by 
\begin{eqnarray}
 |\phi_{pa,out}\rangle &=& |\phi_{pa}, t \to \infty\rangle, \nonumber \\
 |\phi_{pb,in}\rangle &=& |\phi_{pb}, t \to- \infty\rangle,
\end{eqnarray}
 then a typical $\mathcal{S}$-matrix element can be written as
\begin{equation}
\langle|\phi_{pa,out}|\phi_{pb,in}\rangle = \langle|\phi_{pa}|\mathcal{S}^{\dagger}\mathcal{S}|\phi_{pb}\rangle.
\end{equation}
As these charges commute with the Hamiltonian, the time evolution of any physical state will also
also be annihilated by them. 
This implies that the states $\mathcal{S}|\phi_{pb}\rangle$ must be a linear combination of states physical states. 
 So we can write 
\begin{equation}
\langle|\phi_{pa}|\mathcal{S}^{\dagger}\mathcal{S}|\phi_{pb}\rangle = \sum_{i}\langle|\phi_{pa}|\mathcal{S}^{\dagger}|\phi_{0,i}\rangle
\langle\phi_{0,i}| \mathcal{S}|\phi_{pb}\rangle.
\end{equation}
Since the full $\mathcal{S}$-matrix is unitary this relation implies that the  $S$-matrix restricted to physical sub-space is also unitarity. 
It may be noted that the nilpotency of these charges was essential for 
proving the unitarity of the resultant theory. However, in
 massive Curci-Ferrari  gauge these charges are not nilpotent,
\begin{eqnarray}
 Q_B^2 |\phi\rangle &\neq& 0, \nonumber \\ 
 \overline{Q}_B^2 |\phi\rangle &\neq& 0,
\end{eqnarray}
so the $\mathcal{S}$-matrix does not factorize in the massive Curci-Ferrari  gauge,
\begin{equation}
 \langle|\phi_{pa}|\mathcal{S}^{\dagger}\mathcal{S}|\phi_{pb}\rangle \neq \sum_{i}\langle|\phi_{pa}|\mathcal{S}^{\dagger}|\phi_{0,i}\rangle
\langle\phi_{0,i}| \mathcal{S}|\phi_{pb}\rangle,
\end{equation}
and thus  the resultant theory is not unitarity.
However, as the nilpotency is restored 
in the zero mass limit, the unitarity is 
also restored in the zero mass limit. 
 
\section{Conclusion}
In this paper we have analysed a deformed ABJ theory in $\mathcal{N} =3$ harmonic superspace. 
The  classical Lagrangian density 
was  represented by the difference of the two 
 Chern-Simons sectors for the left and right gauge groups plus
the Lagrangian density of the matter fields   which was minimally coupled to the gauge
superfields. No explicit superfield potential was needed in the action. 
We analysed the BRST and 
the anti-BRST symmetries of this model in various gauges. The sum of the ghost term and the gauge fixing term was expressed 
as a combination of a  total BRST and a total anti-BRST variation, in Landau and  Curci-Ferrari gauges.

It will also be interesting to generalized this work to include boundaries. 
The existence of boundaries can have a lot of applications in condensed matter physics, 
due to the existence of edge currents \cite{qhe1}-\cite{qhe2}. 
There is a well-known connection between 
Chern-Simons
theories on a three dimensional manifold and the two dimensional conformal field theories on its boundaries 
 \cite{qft}. 
For pure Chern-Simons theory with suitable boundary conditions, a component of
the gauge field, say $A_0$, appears linearly in the action and so can be
integrated out, imposing the constraint $F_{12} = 0$ \cite{MooreSeiberg}-\cite{EMSS}. This constraint can be
solved explicitly  resulting in  a two dimensional WZW model. 
Even though the  ABJM is not  topological, 
it is still  conformal.  So, in presence of a boundary the ABJM action also gets modified. 
In this  new modified ABJM action  the 
Chern-Simons gauge potential is coupled to a boundary WZW model. This  boundary action reproduces the
pure WZW action when starting from a pure Chern-Simon action. Thus, it 
  is gauge invariant even in presence of a boundary. This has been done for the ABJM theory in 
$\mathcal{N} =1$ superspace \cite{mfa}. This has also been done for the BLG theory in $\mathcal{N} =1$ superspace \cite{mfb}. 

Just like strings can 
end on D-branes in string theory, M2-branes can end on  M5-branes, M9-branes or gravitational waves in M-theory \cite{BCIntMem}.
So, M2-branes in M-theory are analogous to string in string theory. 
Furthermore, just like various  background fluxes can cause various deformations in string theory, the presence of a
background flux can also cause deformation in the M-theory \cite{ChuSehmbi}. Thus, the open        
M2-brane action can be studied to learn about M5-brane action.
The BLG model has been used to motivate a novel quantum geometry on the M5-brane world-volume, 
by analysing    a system of multiple M2-branes ending on a M5-brane with a constant $C$-field 
\cite{d12}. The
 the BLG action with Nambu-Poisson 3-bracket has also been identified with  a M5-brane action, with a 
large worldvolume $C$-field  \cite{M5BLG}. Furthermore, 
 by analysing the action for a single open M2-branes, 
a  non-commutative deformation of string theory on the M5-brane worldvolume has been studied
\cite{NCS1}-\cite{21}. It will also be interesting to analyse these results 
for  the ABJ theory in harmonic superspace.

\end{document}